\documentclass[a4paper,11pt]{article}
\usepackage{pos}
\usepackage{lineno}

\newcommand{\kt}{k_{\text{T}}}
\newcommand{\mt}{m_{\text{T}}}
\newcommand{\qinv}{q_{\text{inv}}}
\newcommand{\Rinv}{R_{\text{inv}}}
\newcommand{\Ntrk}{N_{\text{tracks}}}
\newcommand{\Npart}{N_{\text{part}}}
\newcommand{\TeV}{\text{TeV}}
\newcommand{\GeV}{\text{GeV}}
\newcommand{\PbPb}{\text{PbPb}}

\newcommand{\KZs}{K^0_{\text{s}}}

\title{Recent results on femtoscopic correlations with the CMS experiment}
\ShortTitle{Femtoscopic correlations with the CMS experiment}

\manuallySeparateAuthors
\author*[a,b]{Cesar A. Bernardes}
\author{(on behalf of the CMS Collaboration)}

\affiliation[a]{Instituto de F\'{i}sica, Universidade Federal do Rio Grande do Sul (UFRGS),\\
  Av. Bento Gonçalves 9500, Porto Alegre, Brasil}

\affiliation[b]{Instituto de F\'{i}sica Te\'{o}rica, Universidade Estadual Paulista (UNESP),\\
R. Dr. Bento Teobaldo Ferraz 271, S\~{a}o Paulo, Brasil}

\emailAdd{cesar.augusto.bernardes@cern.ch}

\abstract{The study of femtoscopic correlations in high-energy collisions is a powerful tool to investigate the space-time structure of the particle emitting region formed in such collisions, as well as to probe interactions that the involved particles may undergo after being emitted. An overview of the recent results from the CMS experiment at the LHC on the two-particle femtoscopic correlations measurements using charged particles and identified hadrons in pp and PbPb collisions is presented. In general, the femtoscopic parameters are obtained assuming a Gaussian or an exponential shape to describe the emitting source distribution. In some cases, however, the generalized Gaussian, i.e., the symmetric alpha-stable L\'evy distribution, is favored to describe the source. Some of the measurements allow to extract the parameters of the strong interaction felt by hadrons using their femtoscopic correlations. The studies are performed in a wide range of the pair average transverse momentum (or average transverse mass) and charged particle multiplicities. In addition, prospects for future physics results using the CMS experiment are also discussed.}

\FullConference{%
  41st International Conference on High Energy physics - ICHEP2022\\
  6-13 July, 2022\\
  Bologna, Italy
}

\begin{document}
\maketitle

\section{Introduction}

Femtoscopic correlations are a powerful tool to investigate the spatiotemporal scale of the particle 
emitting source and the final state interactions (e.g. Coulomb and strong interactions) 
among the hadrons produced in hadronic and heavy-ion collisions~\cite{ReviewHBT1}. In this proceedings, results on the two-particle femtoscopic correlations 
are presented using proton-proton (pp) and lead-lead (PbPb) collision data from the CMS experiment at the LHC~\cite{CMSdetector}. 

Femtoscopic correlations using unidentified charged particles are measured in pp collisions with a center of mass energy of $13~\TeV$, analyzing events going up to about $250$ reconstructed charged hadrons~\cite{femtoCMSpp13TeV}. These measurements explore very high particle multiplicity events, where the phenomena of collectivity (similar to the ones from heavy-ion collisions) were observed~\cite{CollectivityInppReview}. The studies were performed using the variable: $\qinv^2 = -q^{\mu}q_{\mu} = -(k_1 - k_2)^2$, where $k_i$ refers to the four-momentum
of each particle of the pair. The correlation function, also known as single ratio, is extracted by dividing the 
$\qinv$ distribution from same-sign particle pairs taken from same event, with a $\qinv$ distribution using pairs from mixed events. 
By fitting the correlation function\footnote{Is important to mention that, in general, methods to subtract non-femtoscopic contributions to the correlation function (mainly from mini-jets, resonances, momentum conservation, flow, etc...) should be applied.} assuming a particle emitting 
source shape (in the case an exponential function), the following parameters are measured: the 
correlation strength ($\lambda$) and the source radius in one dimension (also called the length of 
homogeneity, here denoted as $\Rinv$, or simply $R$). In order to study dynamical behavior of such parameters, 
they are measured as functions of the event multiplicity ($\Ntrk$) as well as 
average pair momentum projection in transverse plane, $\kt = \frac{1}{2}|\vec{p}_{T,1} + \vec{p}_{T,2}|$, and mass, $\mt = \sqrt{m^2_{\pi}+\kt^2}$, where all the hadrons are assumed to be pions. 

In heavy-ion collisions, the Gaussian and exponential functions are usually considered for fitting the correlation function. A generalized form, called L\'{e}vy alpha-stable distribution, has also been considered~\cite{LevyFunc}. This function provided a precise description of the correlation function of the data from the BNL RHIC facility in gold-gold collisions at $200~\GeV$~\cite{PhenixPhysRevC.97.064911}, being characterized by an additional parameter called L\'{e}vy exponent ($\alpha$). The results presented in this proceedings verify the applicability of the L\'{e}vy function at the LHC energies in PbPb collisions, and also investigate if the parameter $\Rinv$ behaves as a particle emitting source size as a function of particle multiplicity (or the number of participants nucleons in the collision, $\Npart$), $\kt$ or $\mt$. Usually the behavior of such parameters as a function of $\Npart$, $\kt$ or $\mt$ can be described by hydrodynamic models assuming a Gaussian shape for the particle emitting source~\cite{PAS-HIN-21-011}.

Femtoscopic correlations using $\KZs$ mesons and $\Lambda$ baryons (called $V^0$ particles) have the advantage to be not 
affected by Coulomb final state interactions. In addition to the 
quantum statistics phenomena with identical particle correlations, the $V^0$ particle correlations are sensitive to 
strong final state effects, which are very important for the understanding of the strong interaction in the non-perturbative scenario. Such measurements have also applicability in the study of the composition of the core of neutron stars~\cite{PAS-HIN-21-006}. 
The results presented in this proceedings consider the Lednicky--Lyuboshits (LL) model~\cite{LLModel} 
to described the strong interaction hadron-hadron scattering parameters. 

\section{Results}

\begin{figure}[!ht]
    \centering
    \begin{minipage}{.45\textwidth}
        \centering
        \includegraphics[width=\linewidth]{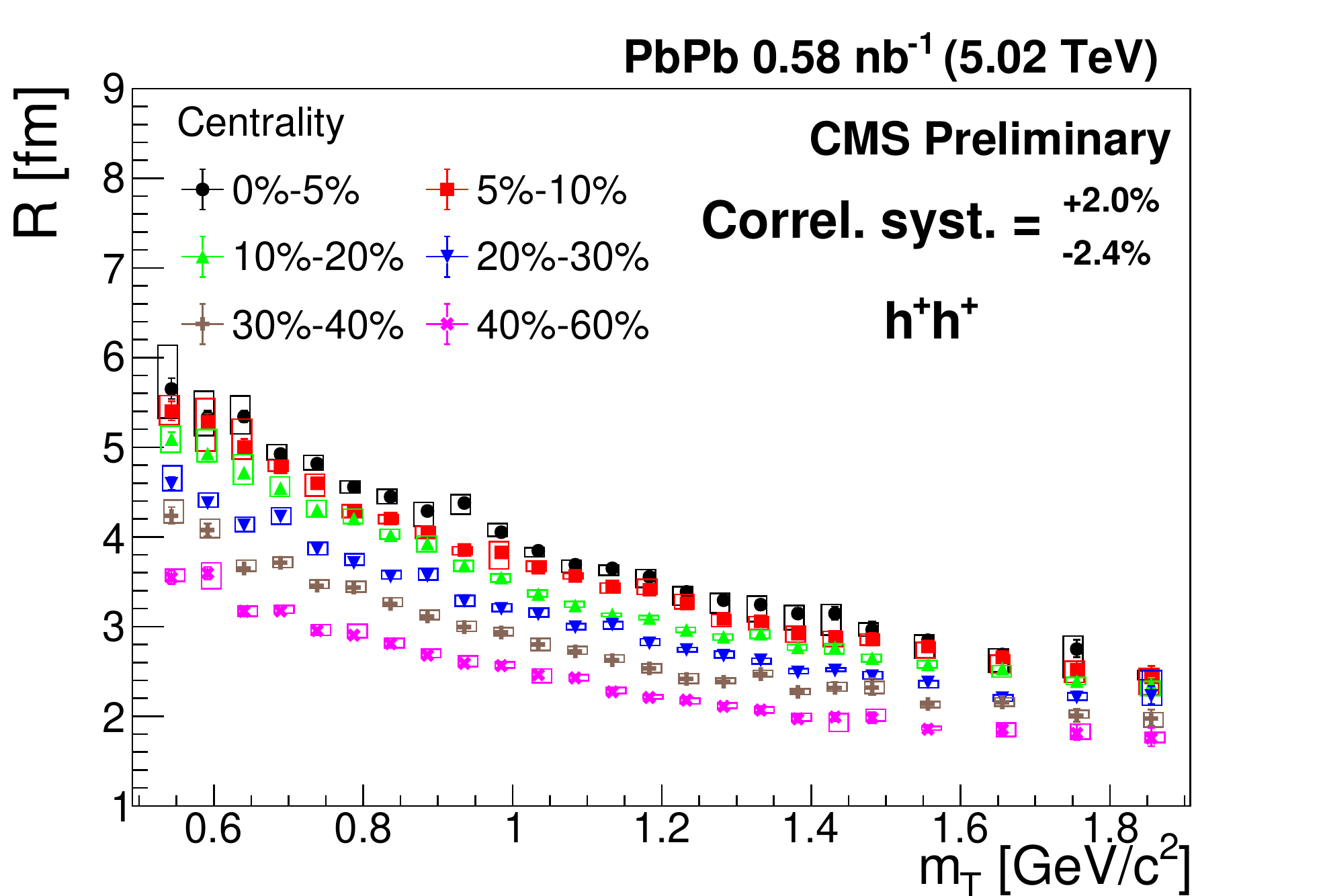}
        \vspace{0.07cm}
    \end{minipage}%
    \begin{minipage}{0.35\textwidth}
        \centering
        \includegraphics[width=\linewidth]{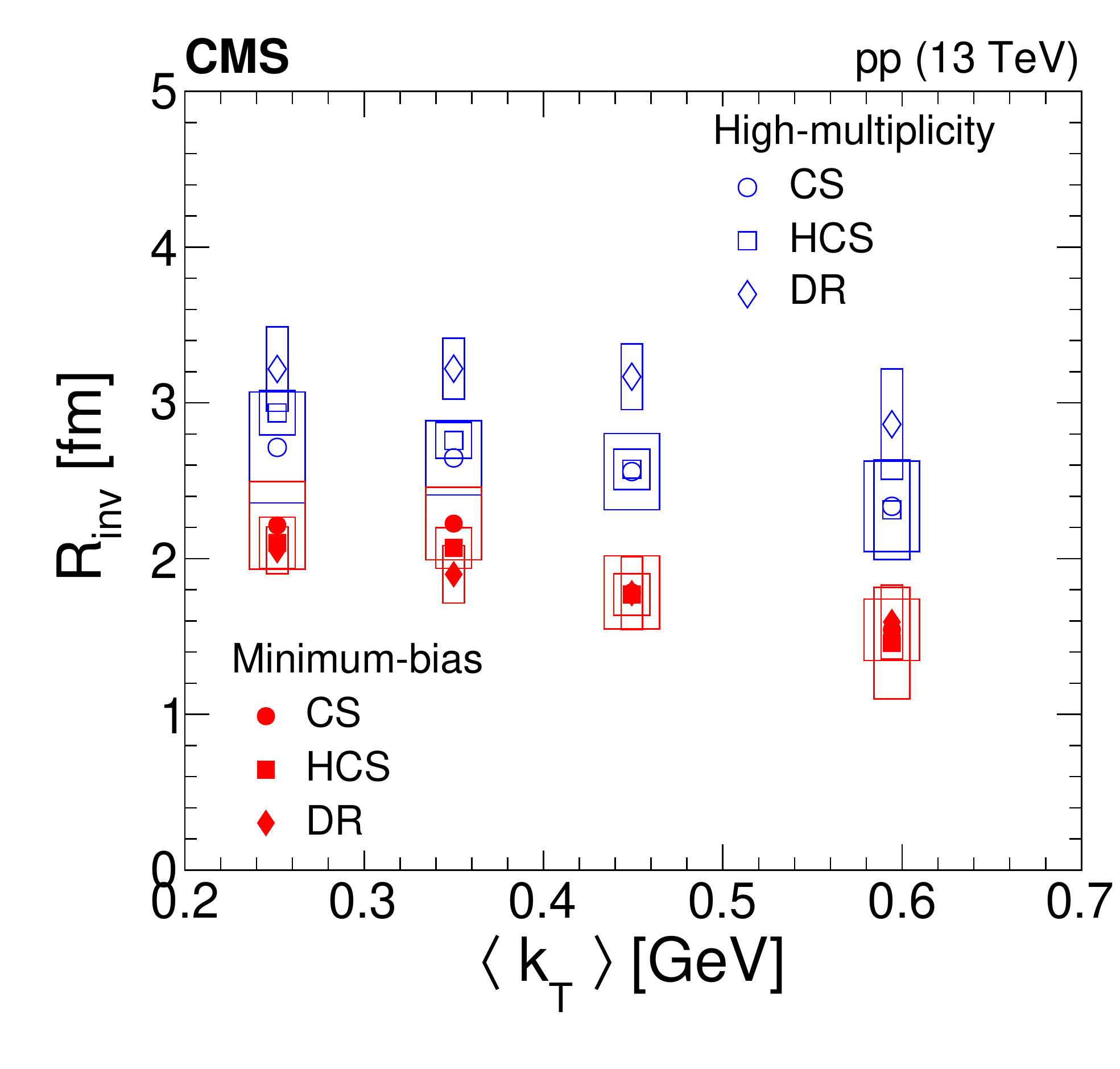}
    \end{minipage}
    \begin{minipage}{.45\textwidth}
        \centering
        \includegraphics[width=\linewidth]{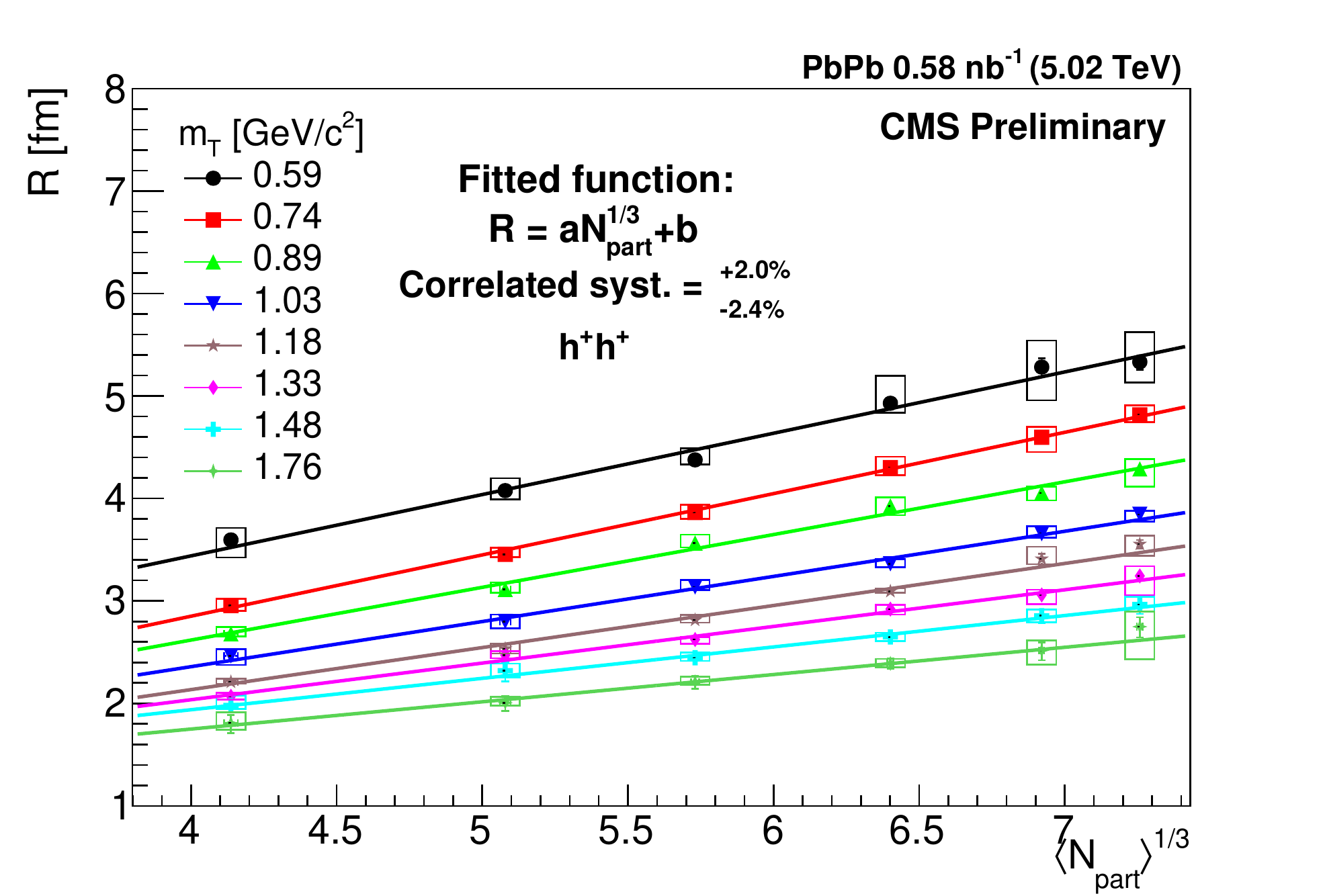}
        \vspace{0.06cm}
    \end{minipage}%
    \begin{minipage}{0.35\textwidth}
        \centering
        \includegraphics[width=\linewidth]{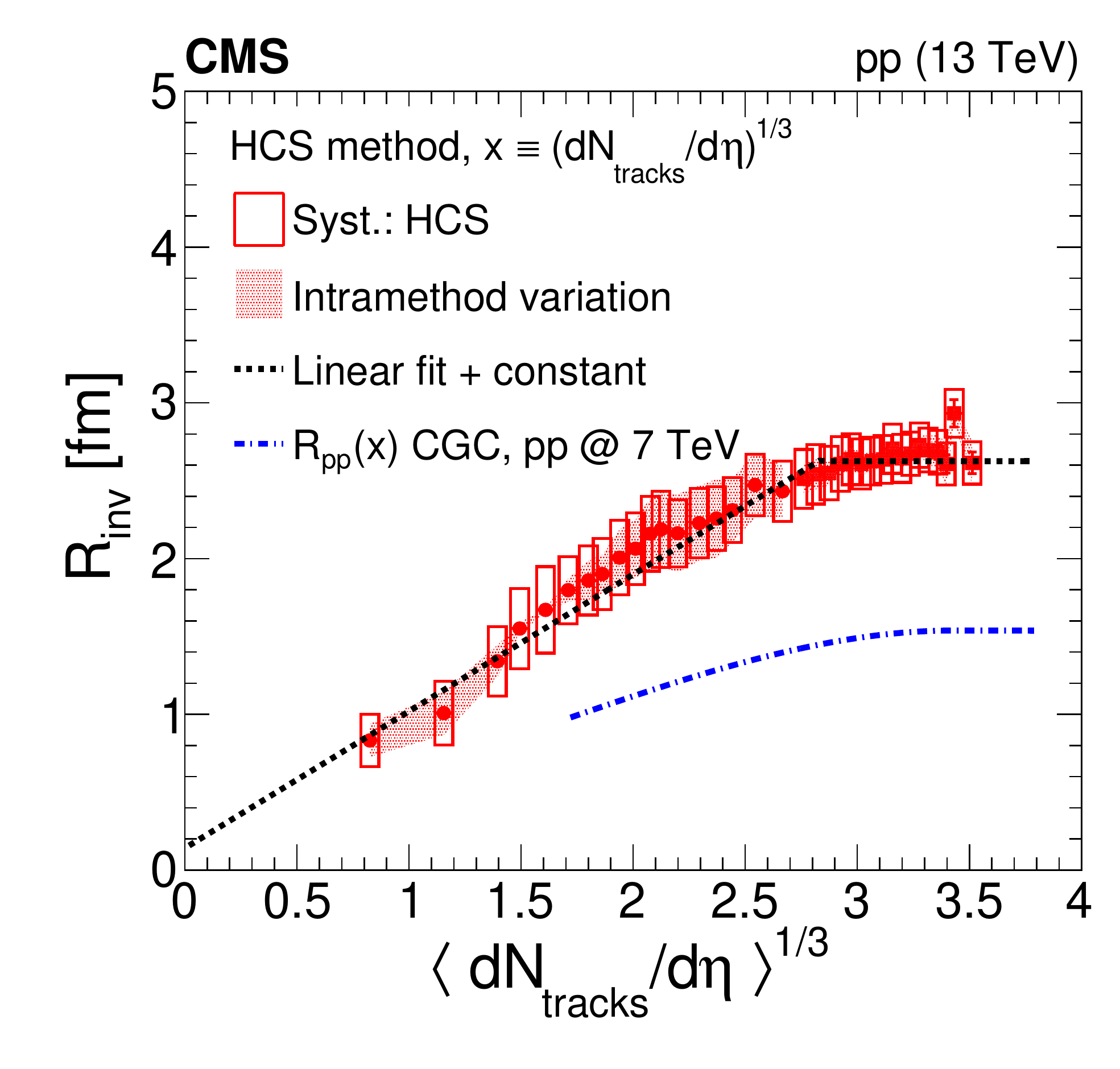}
    \end{minipage}
    \vspace{-0.5cm}
    \caption{(Upper) $\Rinv$ as a function of $\mt$ and $\kt$. (Lower) $\Rinv$ as a function of $\Npart^{1/3}$ and $(d\Ntrk/d\eta)^{1/3}$. Left plots are from $\PbPb$ collisions, while right plots are from pp collisions~\cite{PAS-HIN-21-011, femtoCMSpp13TeV}.}
    \label{fig:fig1}
\end{figure}

Figure~\ref{fig:fig1} shows the measurements of $\Rinv$ as a function of $\mt$ and $\Npart^{1/3}$ for PbPb and 
of $\kt$ and $(d\Ntrk/d\eta)^{1/3}$ for pp collisions. In PbPb, the source scale follows a very similar trend 
as observed when using the Gaussian shape~\cite{PAS-HIN-21-011}, i.e., decreases as a function of $\mt$ and increases linearly as a function of $\Npart^{1/3}$. The measurements as a function of $\mt$ are performed for different centrality classes, which is a measure of the overlap of the two Pb nuclei. In pp collisions, similar trends as compared to PbPb are observed, except for that at higher values of $(d\Ntrk/d\eta)^{1/3}$, where the data could also be consistent with a saturation in the magnitude of $\Rinv$, a feature to be understood. Notice that a very similar trend was also observed in the ATLAS measurements~\cite{ATLAS:2022wvk}.


\begin{figure}[!ht]
    \centering
    \begin{minipage}{.45\textwidth}
        \centering
        \includegraphics[width=\linewidth]{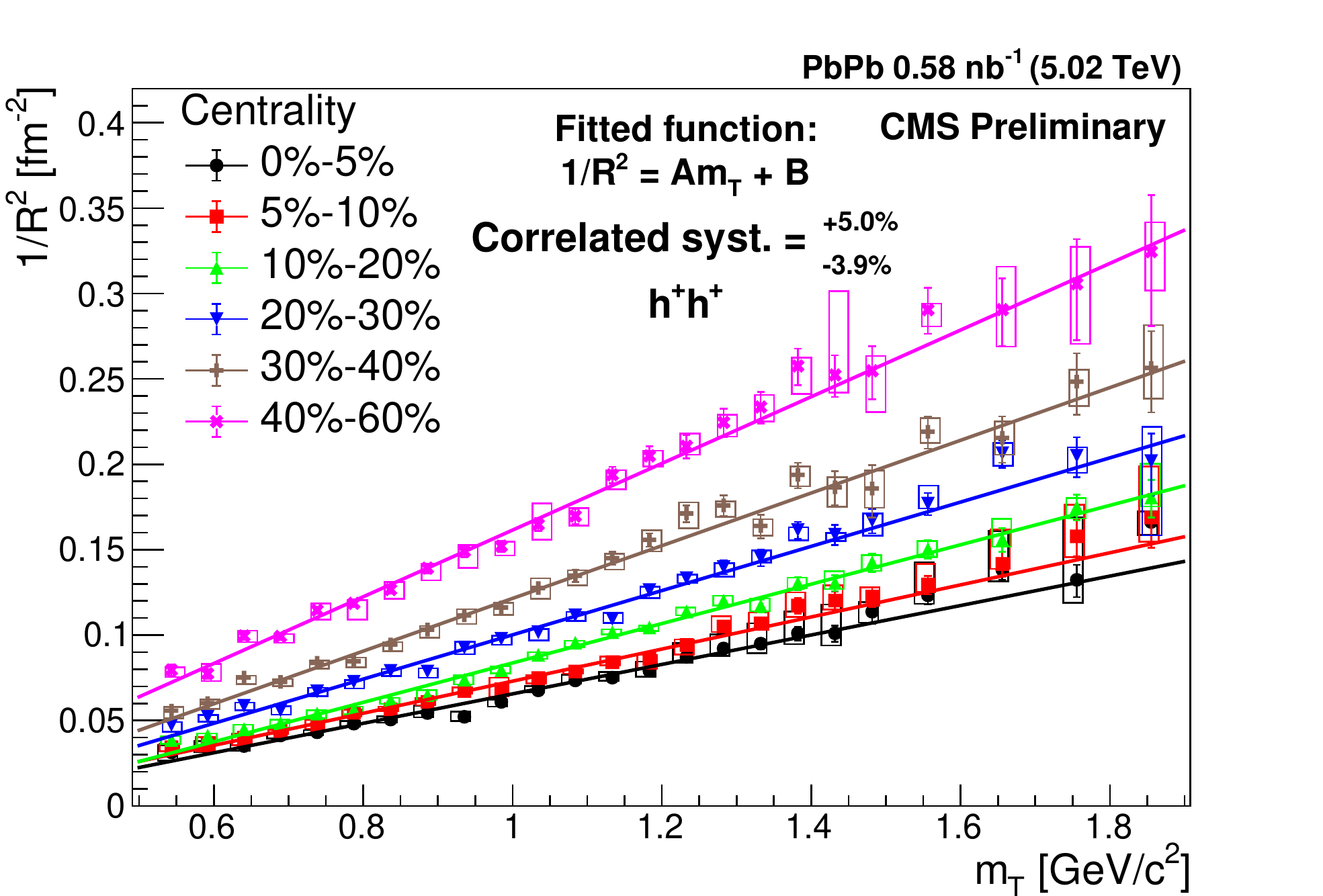}
        \vspace{0.06cm}
    \end{minipage}%
    \begin{minipage}{0.35\textwidth}
        \centering
        \includegraphics[width=\linewidth]{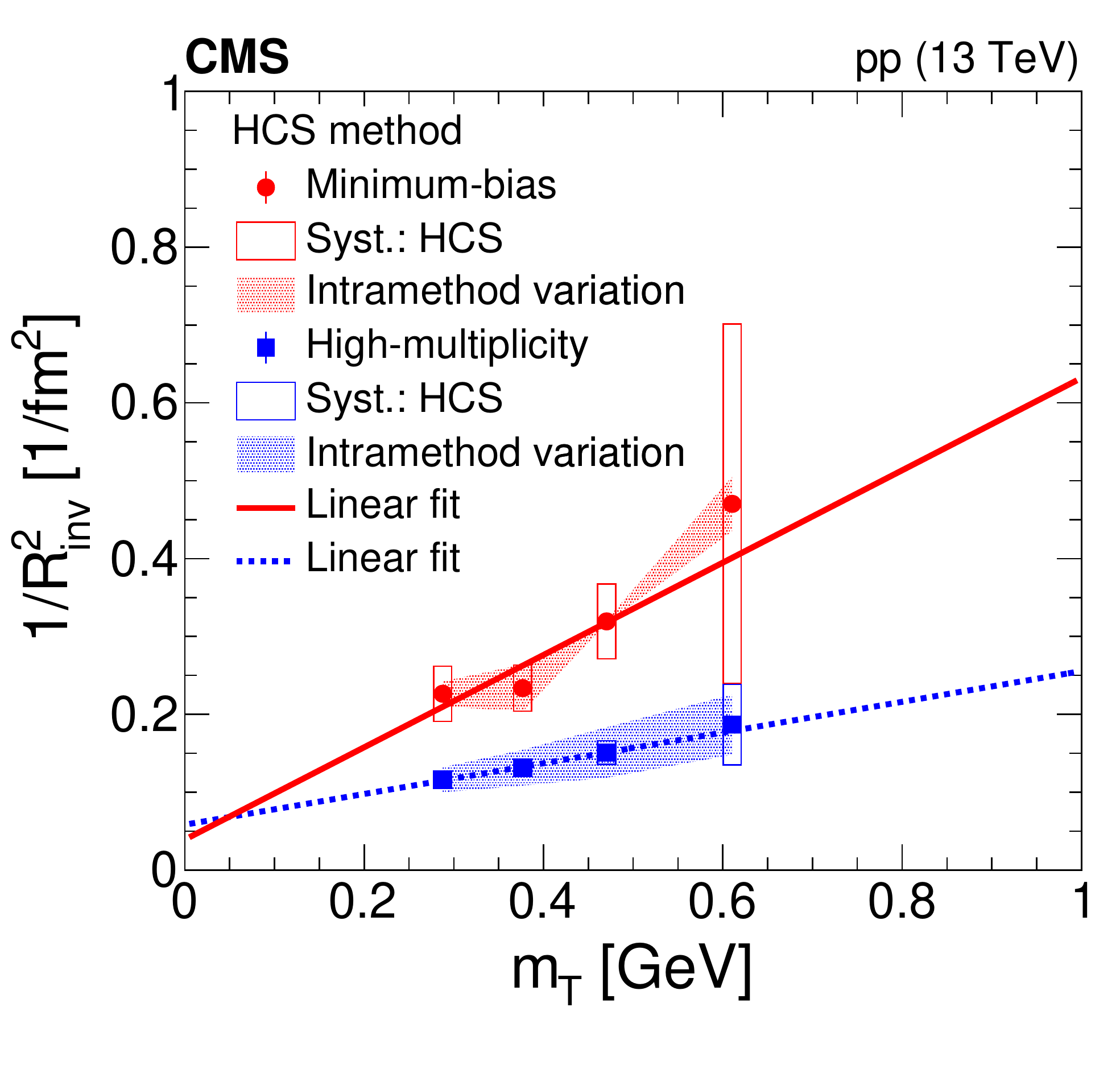}
    \end{minipage}
    \vspace{-0.6cm}
    \caption{$1/\Rinv^2$ as a function of $\mt$ for PbPb collisions (left) and pp collisions (right)~\cite{PAS-HIN-21-011, femtoCMSpp13TeV}.}
    \label{fig:fig2}
\end{figure}

\begin{figure}[!ht]
    \centering
    \includegraphics[width=0.45\linewidth]{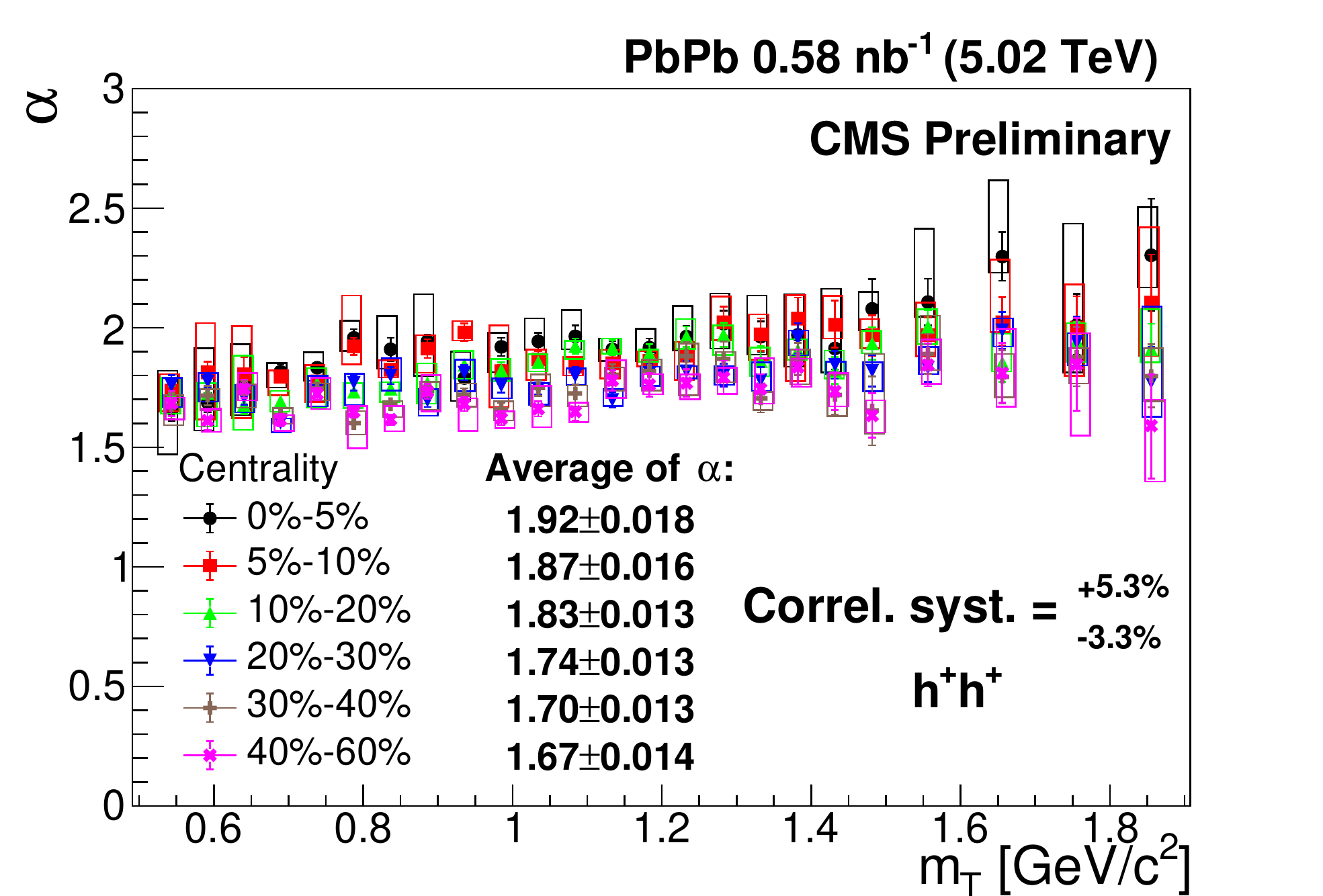}
    \includegraphics[width=0.45\linewidth]{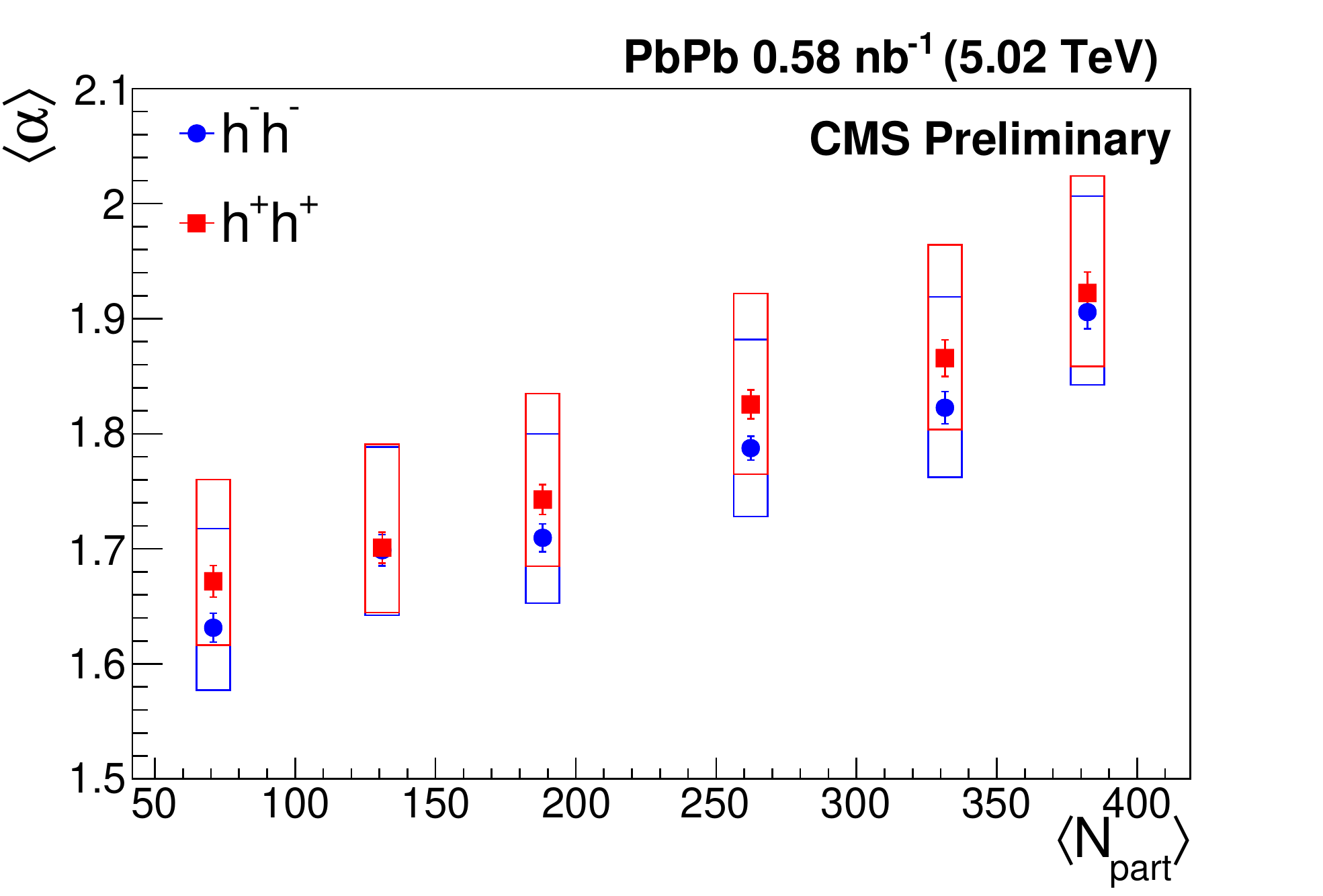}
    \caption{L\'{e}vy $\alpha$ parameter as a function of $\mt$ and its average value as a function of $\Npart$~\cite{PAS-HIN-21-011}.}
    \label{fig:fig3}
\end{figure}

\begin{figure}[!ht]
    \centering
    \begin{minipage}{.45\textwidth}
        \centering
        \includegraphics[width=\linewidth]{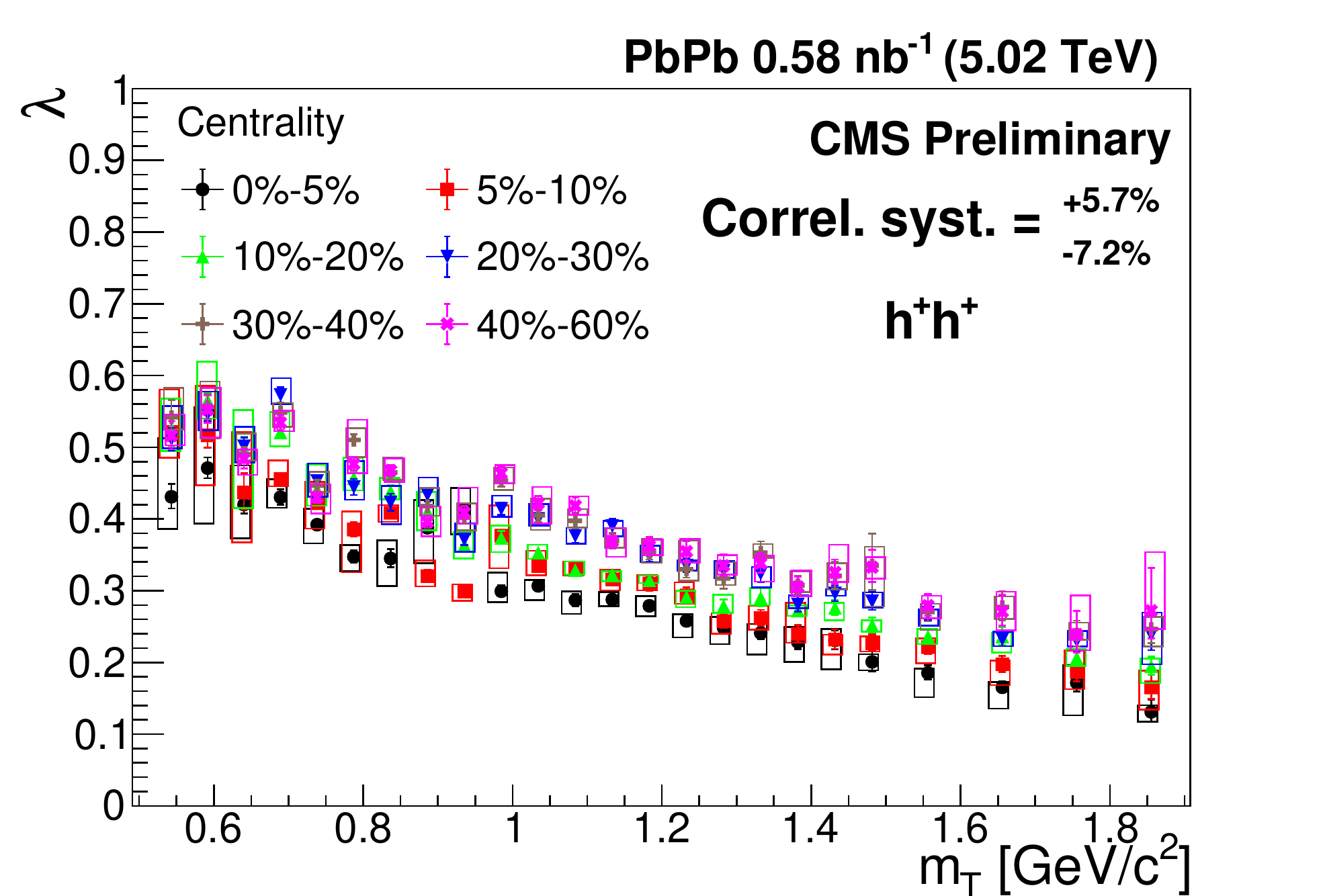}
        \vspace{0.06cm}
    \end{minipage}%
    \begin{minipage}{0.35\textwidth}
        \centering
        \includegraphics[width=\linewidth]{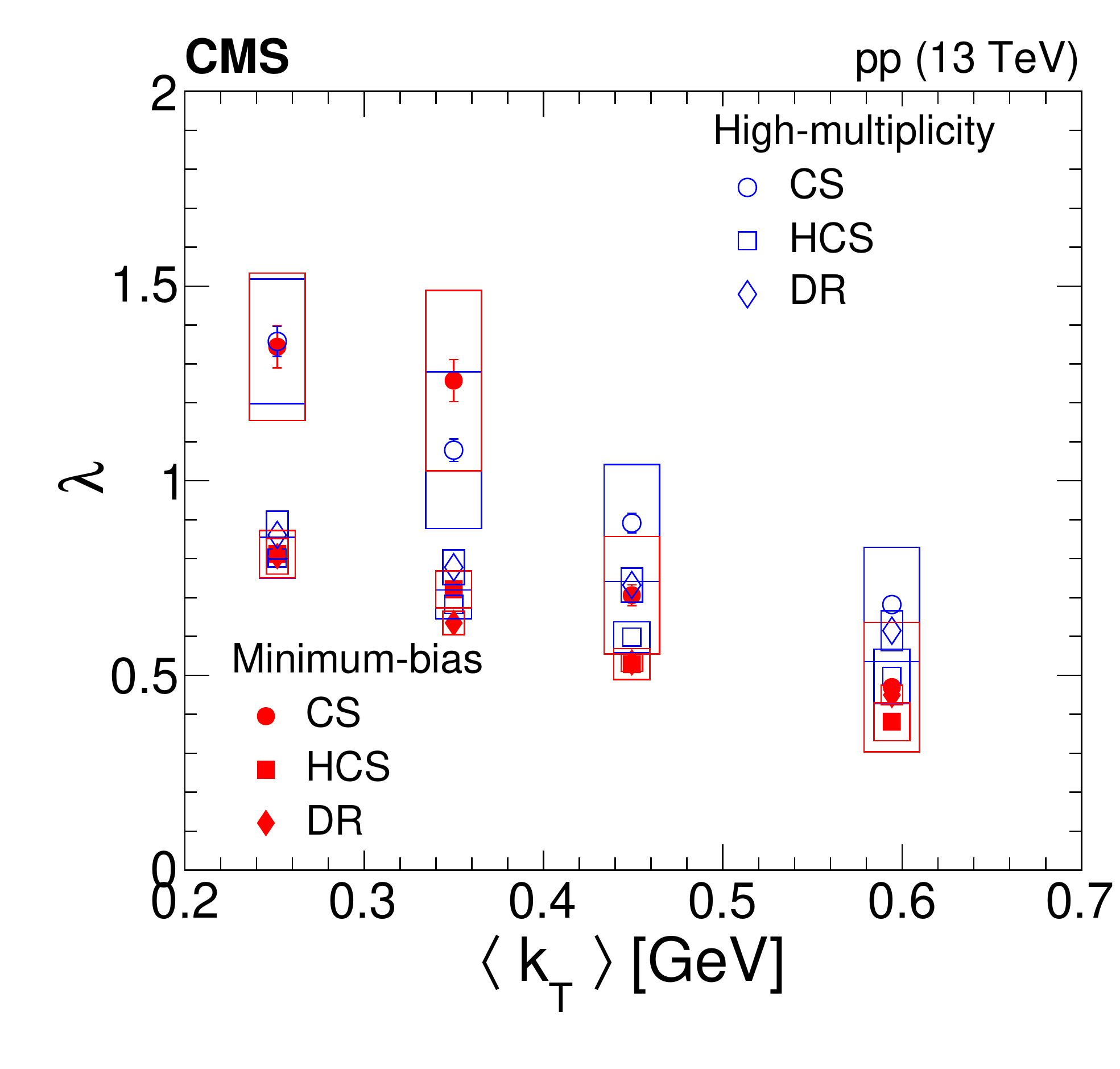}
    \end{minipage}
    \vspace{-0.6cm}
    \caption{Correlation strength parameter as a function of $\mt$ and $\kt$ for PbPb and pp collisions, respectively~\cite{PAS-HIN-21-011, femtoCMSpp13TeV}.}
    \label{fig:fig4}
\end{figure}

In hydrodynamic calculations~\cite{Z.Phys.C.39.1988.69, Phys.Rev.C54.1996.1390}, considering a Gaussian source shape, it is possible to find a relation between $1/\Rinv^2$ and $\mt$. A linear relation is expected, with the slope connected to radial flow (larger slope corresponds to larger radial flow), in addition, the extrapolation to $\mt \rightarrow 0$ is related to the source size at the kinetic freeze-out, called geometrical size. Figure~\ref{fig:fig2} shows these relations for PbPb and pp collisions. For PbPb, using L\'{e}vy function to parameterize the emission source, the behavior on the slope is observed, but the extrapolation of $\mt \rightarrow 0$ gives a negative value, which is not trivial to interpret~\cite{PAS-HIN-21-011}. Figure~\ref{fig:fig3} shows that the $\alpha$ parameter has almost no dependence as a function of $\mt$ variable and a slight dependence on $\Npart$, getting closer to 2 (Gaussian function) for higher values of $\Npart$. A similar behavior for pairs with positive and negative electric charge is seen. As observe in Fig.~\ref{fig:fig4}, the correlation strength decreases with $\mt$ and $\kt$ for PbPb and pp collisions, respectively. A possible explanation for this trend is attributed to the lack of particle identification (see Ref.~\cite{PAS-HIN-21-011} for discussions).


\begin{figure}[!ht]
    \centering
    \includegraphics[width=0.43\linewidth]{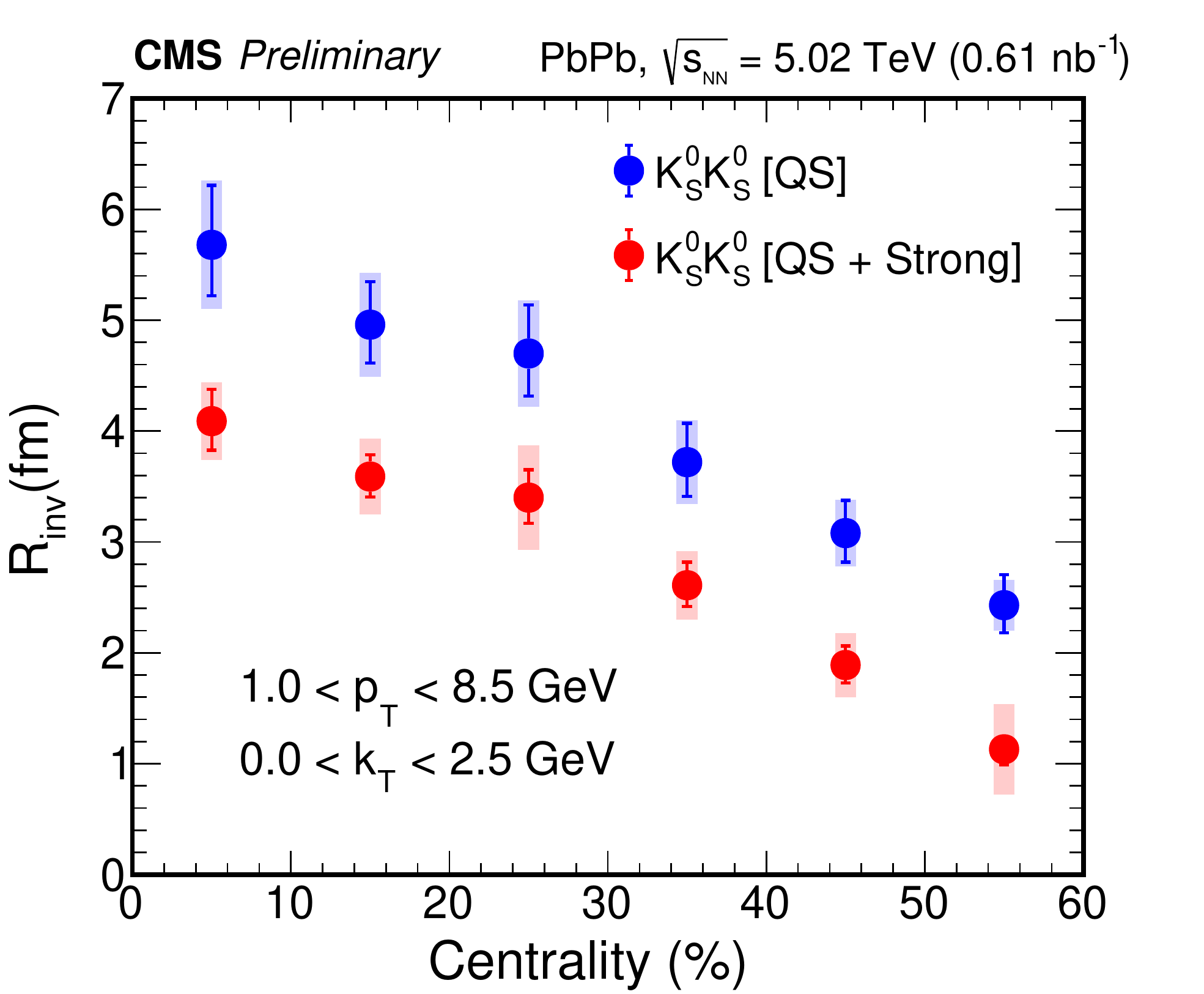}
    \includegraphics[width=0.43\linewidth]{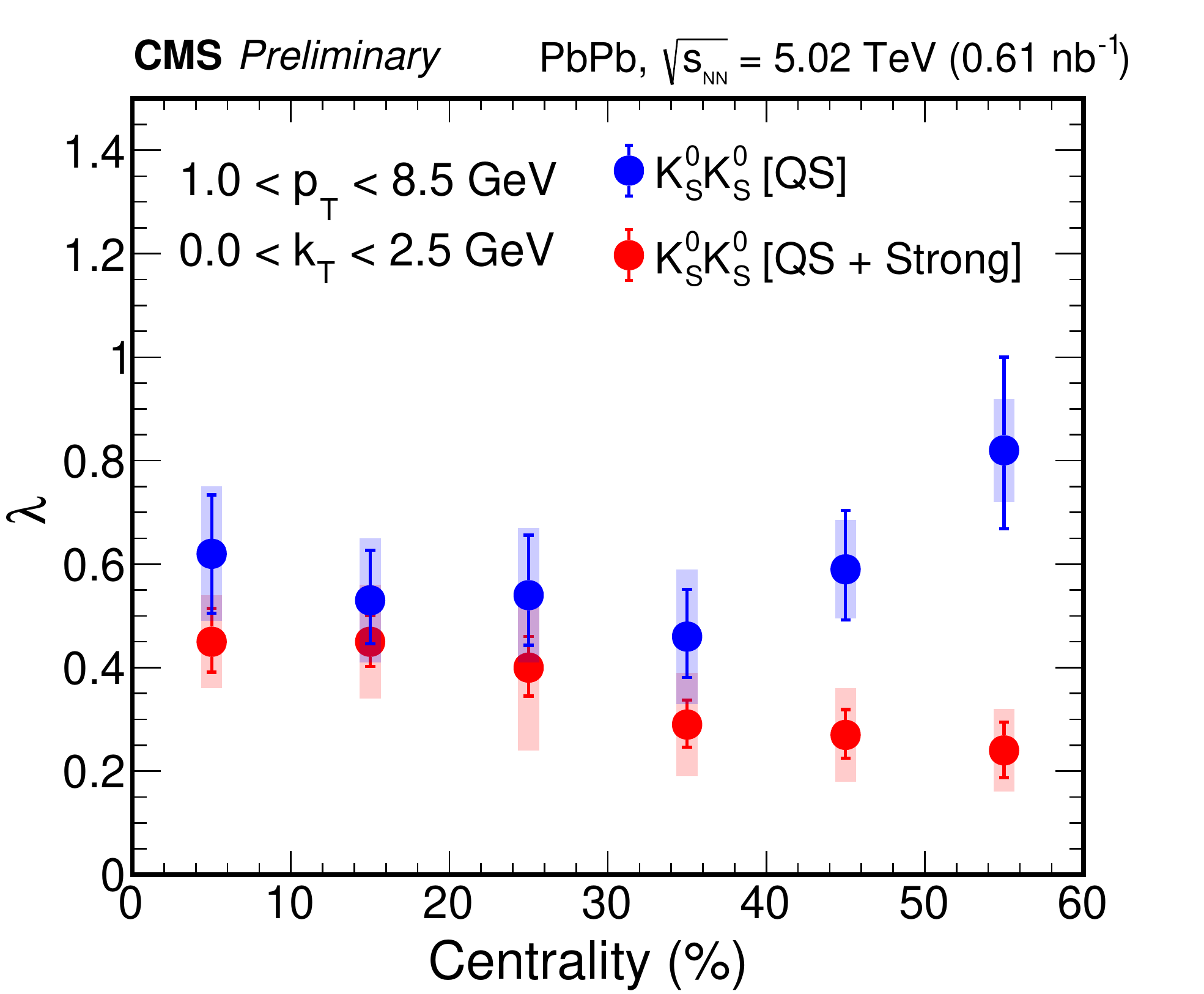}
    \caption{$\Rinv$ and $\lambda$ parameters from $\KZs\,\KZs$ correlations as functions of collision centrality~\cite{PAS-HIN-21-006}.}
    \label{fig:fig6}
\end{figure}

\begin{figure}[!ht]
    \centering
    \includegraphics[width=0.43\linewidth]{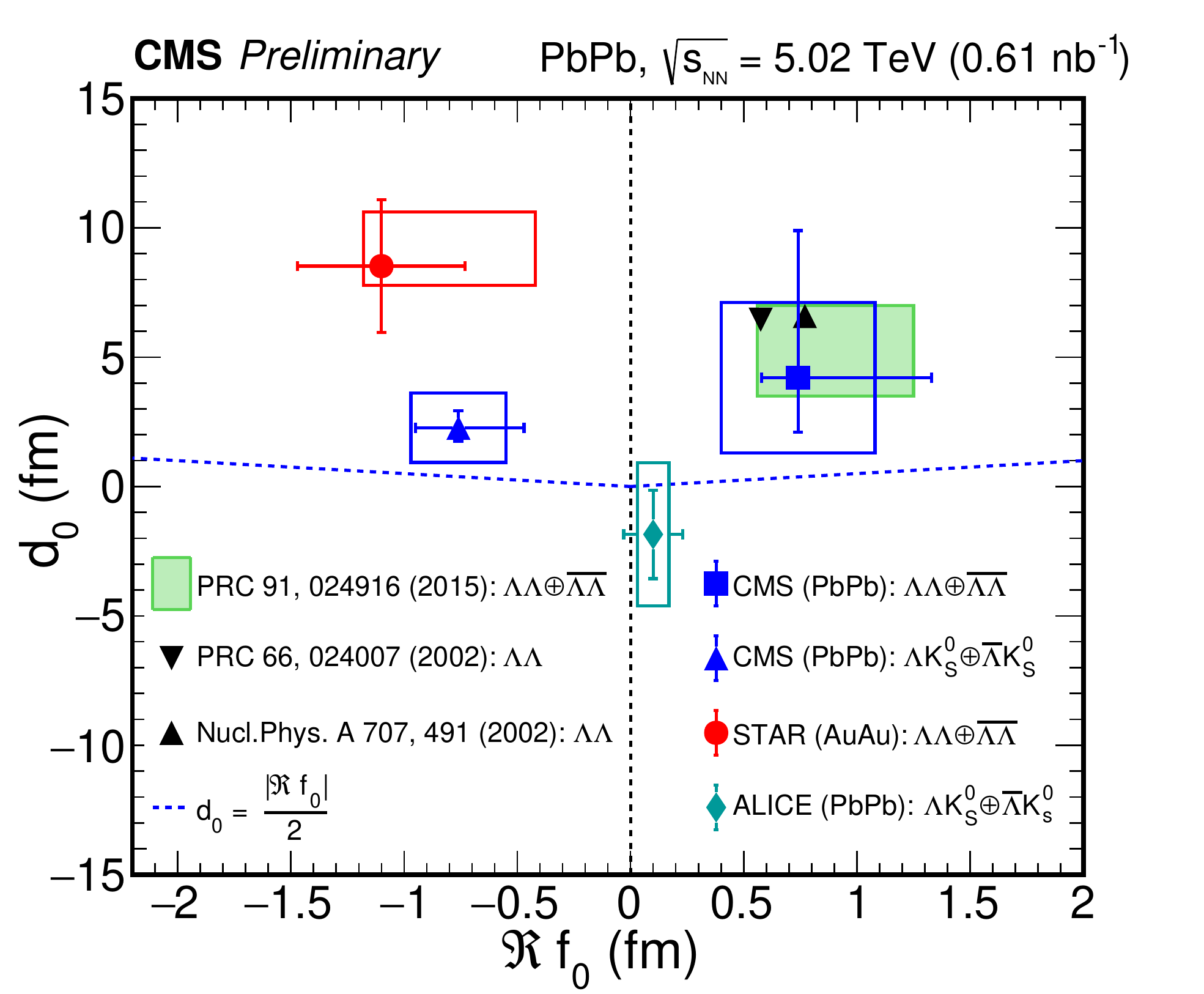}
    \caption{Scattering parameters from the LL model. CMS measurements for $\Lambda\,\Lambda\oplus\overline{\Lambda}\,\overline{\Lambda}$ and $\Lambda\,\KZs\oplus\overline{\Lambda}\,\KZs$ are compared with STAR, ALICE and NAGARA event~\cite{PAS-HIN-21-006}.}
    \label{fig:fig7}
\end{figure}

The $\KZs$ mesons and $\Lambda$ baryons are reconstructed in the $\pi^+\pi^-$ and $p\pi^-$ channels, respectively. 
For the $\KZs\KZs$ correlations, the strong interaction LL model parameters are fixed by using inputs from low energy experiments~\cite{PAS-HIN-21-006}. Figure~\ref{fig:fig6} shows $\Rinv$ and $\lambda$ as functions of collision centrality. 
The $\Rinv$ and $\lambda$ parameters decrease with centrality, similarly as observed in 
many measurements with unidentified and other identified particles. The effect of the strong interaction 
parameterization in the values of $\Rinv$ and $\lambda$ was also tested, showing a considerable difference between both cases. 
The $\Lambda$ baryon correlations are used to measured the scattering parameters of the LL model, such as 
the real part of the scattering length ($\mathcal{R}f_0$) and effective range ($d_0$)~\cite{PAS-HIN-21-006}. 
The cases with $\mathcal{R}f_0 < 0$ are interpreted as an anticorrelation (depletion below unit or a repulsive interaction) and the scenarios with $\mathcal{R}f_0 > 0$ are interpreted as a correlation above unit (attractive interaction). Figure~\ref{fig:fig7} 
shows these parameters for the $\Lambda\,\Lambda\oplus\overline{\Lambda}\,\overline{\Lambda}$ and 
$\Lambda\,\KZs\oplus\overline{\Lambda}\,\KZs$ correlations, indicating an attractive interaction for $\Lambda\,\Lambda\oplus\overline{\Lambda}\,\overline{\Lambda}$ and a repulsive interaction for $\Lambda\,\KZs\oplus\overline{\Lambda}\,\KZs$.
The result for $\Lambda\,\KZs\oplus\overline{\Lambda}\,\KZs$ differs from those reported by the ALICE collaboration~\cite{Phys.Rev.C103.2021.055201} for PbPb collisions at $2.76~\TeV$.


\section{Conclusions}

In summary, femtoscopic correlations for charged hadrons, $\KZs$ mesons, and $\Lambda$ baryons
are measured in proton-proton and lead-lead collisions with the CMS experiment. 
The measured particle emitting source parameters in general show similar behavior between the two colliding systems.
For charged hadrons, in PbPb collisions, a L\'{e}vy alpha-stable distribution is used to describe the correlation 
function, showing a very good description of the data and fitting parameters with dynamical behavior 
compatible with a particle emitting source size. Measurements using the 
$\KZs$ mesons and $\Lambda$ baryons suggest an attractive and a repulsive interaction for $\Lambda\,\Lambda\oplus\overline{\Lambda}\,\overline{\Lambda}$ and $\Lambda\,\KZs\oplus\overline{\Lambda}\,\KZs$, respectively.

\acknowledgments{This material is based upon work supported by FAPESP under Grant No. 2018/01398-1, by FAPERGS Grant No. 22/2551-0000595-0, and by CNPq Grant No. 407174/2021-4.}

\end{document}